# Novel method for handling Ethereum attack

G. Hall, Mansi M., Makrant I.

2019

Abstract

Block-chain world is very dynamic and there is need for strong governance and underlying technology architecture to be robust to face challenges. This paper considers Ethereum, a leading block chain. We deep dive into the nature of this block chain, wherein for software upgrades forks are performed. They types of forks and impact is discussed. A specific Ethereum hack led to a hard fork and focus is provided on understanding the hack and overcoming it from a novel approach. The current model has been unable to handle multiple Ethereum attacks. Thus the current approach is compared against a novel approach providing a security and scaling solution. Here the architecture draws upon combining block-chain layers into operating system level. The approach can have tremendous benefits to block chain world and improve the way decentralized application teams perform. The benefits of the novel architecture is discussed. The approach helps safe guard block chain projects, making them safer and chain agnostic.

## 1   Introduction

Ethereum is a leading distributed block-chain. Ethereum is capable of writing code for block-chain in decentralized manner for applications. Solidity is the programming language used for writing smart contracts. Sharing and exchange of money is possible through applications established Ethereum network. This is trustless system and no middle men are involved. There is no downtime as the nodes do not stop running. Users are protected from censorship and third party influences. Instead of building individual block chains, Ethereum virtual machine can be used to host the block-chain. Nodes are used to connect the block chain. The node usually uses an antenna model for communication within a device or multiple nodes that form a cluster. The blocks in Ethereum are designed in hierarchical structure and so you can match semantic of each block id, transaction id to search for the transactions occurred within blocks. As multiple blocks are connected in ontological structure, the semantic of block-id, transaction-id and account-id can be used.

Ethereum can help build decentralized autonomous organizations known as DAO. There help funding of development for profit and non-profit focus. These are multiple users take part with no leader and this continues till Ethereum governance has faced lots of challenges over a period of time. Once such occurrence was during DAO attack. In order to understand DAO (Decentralized autonomous organization), we describe the types of Ethereum forks.

## 2 Model for handling software update

Forks are a way to bring about software changes. Develoeprs make enhancement to chain inorder to ensure the protocol is continously improving over time. There is need for a process to bring about changes via forks. Core code of protocol should be protected, so that the block-chain databases are protected and compatible. Testing should also be done so that block-chain should not break down. There are two types of Ethereum fork that is hard fork and soft fork.

### 2.1 Soft Fork

There are changes to software configuration part. There are provided by the core development team.

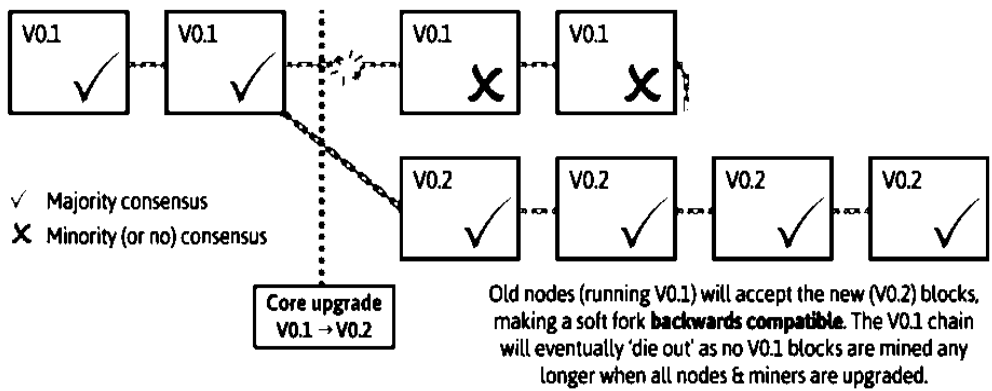

*Fig 1 Soft Fork*

Consider the above figure wherein softfork occurs at a given snapshot. Here we get two chains, the exiting one known as V0.1 and new ones VO.2. The dual forks will stop to exist that is become one when all the node operators fully upgrade. The chain does not actually split as there is consensuses on proposed change is approved by the stakeholders. There is backward compatibility for soft fork. This implies that the old nodes have the capability to process new version.

2.2    Hard Fork

For a radical change to be made to block-chain protocol is enabled through hard fork. Resultantly all the key stakeholders such as the network operator and user have to upgrade to the latest protocol. There is permanent shift from the older version of the chain. Prior version is no more acceptable by latest version.

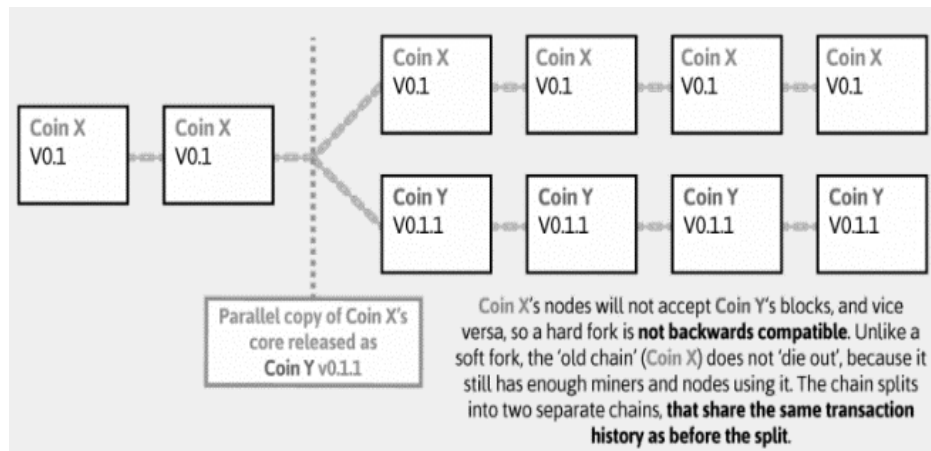

Fig 2 Hard Fork

Consider above figure. There is base coin X having version V0.1. At a given point, a snapshot of hard fork is decided. At that snapshot, we get two chains X and chain Y. These chains are independent chains. The new chains try to compete among each other for survival, node operation, developer and branding. The only thing the common thing these chains share is the transaction history before the snapshot is taken. These chains behave differently having different hashing power, mining difficulty and other metrics. The old version refuse to recognize the new blocks.

There is need to provide important security risks in old version. This radical change can be done through hard fork. In order to add new features or rollback certain transactions, hard fork is needed.

## 3    Decentralized Autonomous Organization

The need to provide decentralized business setup for nonprofit and profit was provided by DAO that stood for decentralized autonomous. This was setup on Ethereum block chain. There was no formal board members.  The DAO code had a limitation due to which hack took place in the month of June 201. The issue let to loss of funds for DAO token holders and resultantly Ethereum members. Block chain is characterized by non rollback that transactions once placed, should be honored. The difference of opinion among Ethereum community led to a hard fork of Ethereum known as Ethereum Classic. These are two separate and independent chains.

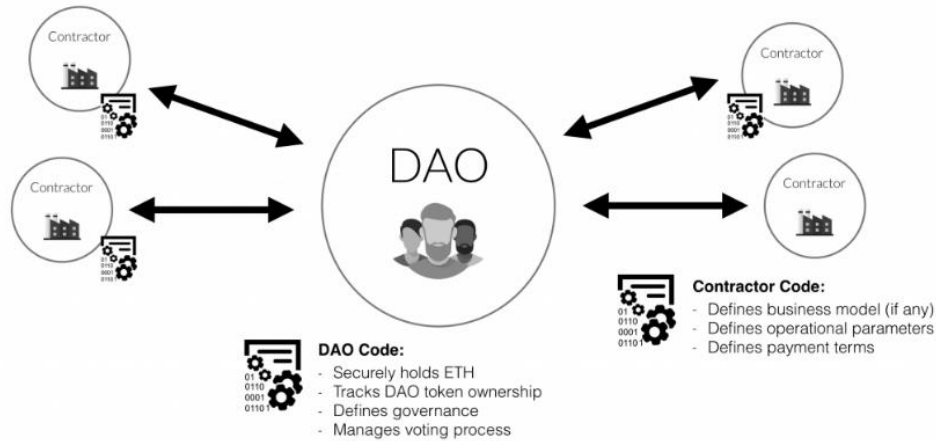

*Fig 3 DAO Architecture*

Ethereum discussion to rollback took block chain community by shock as innate state of block chain should be irreversible.

Potential Resolution

The DAO code bug occurred in the layer 3 that deals with the contract code. This comes under cyber security and system programming.

The token interface creates a map data structure in which all map data structures are mapped. The map helps to track the specified addresses where tokens are sent. The total supply is the variable which tells us what is available in the contract. The balance will tell us the balance of token contained in that particular address. Transfer is to send token from one address to another address. There is a TransferFrom wherein transfer is performed on behalf of a person's approval.

Provided below is the TokenInterface code that follows above process.

contract TokenInterface {

mapping1 (address1 => uint256) balances;

mapping1 (address1 => mapping1 (address1 => uint256)) allowed;

uint256 public totalSupply;

function balanceOf(address1 _owner) constant returns (uint256 balance);

function transfer(address1 _to, uint256 _amount) returns (bool success);

function transferFrom(address1 _from, address1 _to, uint256 _amount) returns (bool success);

function approve(address1 _spender, uint256 _amount) returns (bool success);

function allowance(address1 _owner, address1 _spender) constant returns (uint256 remaining);

event Transfer(address1 indexed _from, address1 indexed _to, uint256 _amount);

event Approval(address1 indexed _owner, address1 indexed _spender, uint256 _amount);

}

# 4   Recommended novel architecture to handle DAO

The traditional model that is existing block-chain have base layer as layer one. Layer two and layer three are different for each block-chain. This implies the process is currently not chain agnostic.

## 4.1   Traditional Model

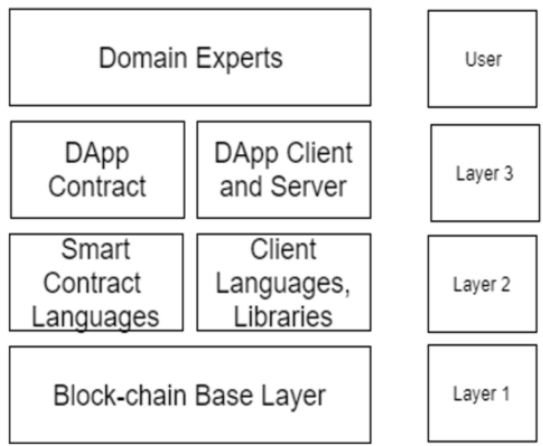

Fig.1 Traditional model

Consider above figure. Ethereum block-chain has base layer 1. This is foundation layer. Layer 2 comprises of Smart Contract Languages. Layer 2 also includes client languages and libraries. Layer 3 comprises of smart contract related stuff. This also includes decentralized client and server model. The topmost layer comprises of end users who are generally domain experts. In above scenario there is heavy dependency of users should know about the development, block-chain domain in addition to their own domain expertise. This leads to lots of overhead and also lack of focus. Similar the developers need to not only focus on their development plus need to learn specific block-chain related development and cryptography related concepts. If tomorrow block-chain changes, developers need to learn a new block-chain language. Developers too would need to understand business domain as technology drives business in this model. Thus there is extreme pressure to hire a large team of experts that need to wear multiple hats.

## 4.2    Recommended Architecture

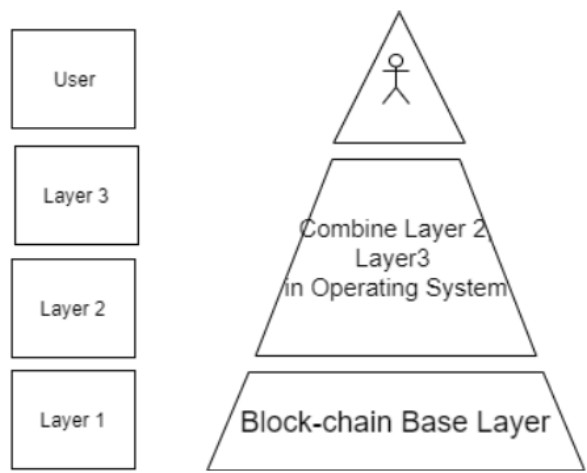

*Fig 4 Operating System -Novel Architecture*

Consider above figure wherein a better approach is provided that involves Layer2 and Layer3 architecture for Operating System. Layer 1 is the traditional block chain layer. Layer 2 and Layer 3 is where the emphasis is wherein new operating system is established for handling cyber security, distributed systems, cryptography and system programming. This would have helped the developers to focus on building decentralized applications. This enables developers to focus on their main role of building features. Code development is standardized and risk free.

Above operating system development would have helped overcome DAO hack, Ethereum hard fork, loss of funds and reputation due to the event. Optimal Operating System can also help inter block chain communication and facilitate more transactions among different chains. This also can help sister and side chain to gain recognition. Thus adoption of Ethereum will increase tremendously. Domain experts can focus exclusively on their area of strength and need not worry about other items. Currently an extremely skilled developer is needed to have block chain knowledge and also industry domain expertise, causing lack of focus which is resolved by this layer2 solution.

## 5    Conclusion

DAO hack led to a hard fork for Ethereum leaving lots of concerns of the block chain, even in present day. Millions of dollars was lost. The multiple risks associated with DAO hack can be addressed by Operating System level focus. This would put the application developers in control, by focussing on their strength while the new Operating System builds the safeguards, needed to keep the block chain safe, secure. Domain experts would focus on their specific expertise, without factoring block chain related nuances. This would help block chain start-ups grow as they gain relevant focus and operate in secure manner as technical risks are addressed at Operating System level. Block chain is continuously evolving and changes can cause disruption. Hence a block chain agnostic approach by Operating System level focus helps to address this. The extremely stressing scaling needs too are addressed by the approach.